\documentclass[prc,floatfix,groupedaddress,nofootinbib,showpacs,preprintnumbers,
amsmath,amssymb,amsfonts,superscriptaddress,widetable] {revtex4}
\usepackage{graphicx}% Include figure files
\usepackage{dcolumn}%Align table columns on decimal point
\usepackage{mathrsfs}
\usepackage{bm}% bold math

\usepackage{graphicx}% Include figure files
\usepackage{dcolumn}% Align table columns on decimal point
\usepackage{bm}% bold math
\usepackage[usenames]{color}
\begin{document}

\title{Relativistic effective interaction for nuclei, giant resonances,
         and neutron stars}
\author{F. J. Fattoyev}~\email{ff07@fsu.edu}
\affiliation{Department of Physics, Florida State University,
             Tallahassee, FL 32306}
\author{C.~J.~Horowitz}~\email{horowit@indiana.edu}
\affiliation{Nuclear Theory Center and Dept. of Physics, Indiana
University, Bloomington, IN 47405}
\author{J. Piekarewicz}~\email{jpiekarewicz@fsu.edu}
\affiliation{Department of Physics, Florida State University,
             Tallahassee, FL 32306}
\author{G. Shen}~\email{gshen@indiana.edu}
\affiliation{Nuclear Theory Center and Dept. of Physics, Indiana
University, Bloomington, IN 47405}

\date{\today}
\begin{abstract}
Nuclear effective interactions are useful tools in astrophysical
applications especially if one can guide the extrapolations to the
extremes regions of isospin and density that are required to simulate
dense, neutron-rich systems.  Isospin extrapolations may be
constrained in the laboratory by measuring the neutron skin thickness
of a heavy nucleus, such as $^{208}$Pb. Similarly, future observations
of massive neutron stars will constrain the extrapolations to the
high-density domain. In this contribution we introduce a new
relativistic effective interaction that is simultaneously constrained
by the properties of finite nuclei, their collective excitations, and
neutron-star properties. By adjusting two of the empirical parameters
of the theory, one can efficiently tune the neutron skin thickness of
$^{208}$Pb and the maximum neutron star mass. We illustrate this
procedure in response to the recent interpretation of X-ray
observations by Steiner, Lattimer, and Brown that suggests that the
{\sl FSUGold} effective interaction predicts neutron star radii that
are too large and a maximum stellar mass that is too small.  The new
effective interaction is fitted to a neutron skin thickness in
$^{208}$Pb of only $R_n\!-\!R_p\!=\!0.16$~fm and a moderately large
maximum neutron star mass of 1.94~$M_\odot$.
\end{abstract}
\maketitle

\section{Introduction}
\label{intro}
Nuclear effective interactions provide a compact and efficient
description of both the structure of finite nuclei and the equation of
state---pressure as a function of density and temperature---of nuclear
matter. Effective interactions are often fitted to well determined
nuclear observables, such as binding energies and charge radii of
closed shell nuclei.  However, limiting the calibration procedure to
these few properties leaves important parts of the effective
interaction largely undetermined, particularly the isovector
interaction.  Failure to constrain the isovector interaction affects
the properties of systems with large isospin asymmetry.  In the case
of finite nuclei, perhaps the best example is the neutron radius (or
neutron skin) of heavy nuclei. Indeed, effective interactions
calibrated to the binding energy and charge radii of closed-shell
nuclei often predict significantly different neutron
radii~\cite{Brown:2000,Furnstahl:2001un}. In the case of infinite
matter, it is the density dependence of the symmetry energy that
remains largely unconstrained. These poorly constrained properties
are very important in astrophysical systems with large neutron
asymmetries, such as neutron
stars~\cite{Horowitz:2000xj,Horowitz:2001ya,Horowitz:2002mb,
Steiner:2004fi,Li:2005sr,Lattimer:2006xb}.

In this article we develop an effective interaction suitable for the
calculation of the ground-state properties of finite nuclei, their
collective response, and the structure of neutron stars.
Astrophysical predictions are notoriously difficult as they involve
two large extrapolations away from the familiar landscape of stable
nuclei. Whereas one involves isospin extrapolations from the domain of
stable nuclei to the very neutron-rich systems present in neutron
stars, the other one involves an extrapolation from the density of
normal nuclei to the very high (and very low) densities that are
encountered in neutron stars.  To reduce the uncertainties associated
with these extrapolations we impose two additional constraints---one
experimental and one observational---on the effective interaction.
Specifically, we extend the standard protocol of fitting to the
binding energies and charge radii of finite nuclei by demanding that
the new effective interaction also reproduces: (a) the neutron radius
of ${}^{208}$Pb and (b) the maximum mass of a neutron star.  We have
selected these two quantities over other possible choices because they
can be determined in a model independent manner from experiment and
observation.  In particular, fixing the neutron radius in ${}^{208}$Pb
reduces significantly the uncertainties associated with extrapolations
in isospin.  The Lead Radius experiment (PREx)---which was
successfully commissioned in March of 2010---aims to determine the
neutron radius of $^{208}$Pb in a model independent fashion by using
parity violating electron
scattering~\cite{Michaels:2005,Horowitz:1999fk}.  To reduce the
uncertainty associated with extrapolations to high densities, we have
selected to fix the maximum neutron star mass.  Of course, the maximum
neutron star mass selected for the fit must at least be equal to the
largest---well measured---mass determined from
observation~\cite{Freire:2007sg}.

In principle, other choices are possible to guide these
extrapolations.  For example, laboratory experiments with heavy ions
have played a critical role in probing the nuclear equation of
state. By tuning the energy of the colliding beams and the
neutron-proton asymmetry, heavy-ion collisions probe vast regions of
the phase diagram.  In particular, isospin diffusion data may be used
to constrain the density dependence of the symmetry
energy~\cite{Tsang:2004zz,Tsang:2008fd,Shetty:2005qp}.  This could be
used instead of the neutron radius in ${}^{208}$Pb.  Moreover,
experiments with very energetic heavy ions have compressed nuclear
matter to densities in excess of four times nuclear matter saturation
density~\cite{Danielewicz:2002pu}. The determination of the pressure
of symmetric nuclear matter at these high densities may be used
instead of the maximum neutron star mass. However, not only is the
interpretation of heavy-ion data model dependent, but heavy-ion
systems are much less neutron rich and often much hotter than neutron
stars. This introduces large and uncontrolled uncertainties.

Neutron star radii as a function of mass---the so-called {\sl
Mass-vs-Radius} relationship---provides critical information on the
equation of state of neutron-rich matter at high densities. Yet, the
simultaneous determination of both the mass and radius of a neutron
star is difficult.  However, tremendous advances in X-ray astronomy
have produced large amounts of high quality data that have allowed the
simultaneous determination of both masses, radii, and the underlying
equation of state~\cite{Ozel:2010fw,Steiner:2010fz}.  Unfortunately,
neutron star radii inferred from X-ray observations of the luminosity
and temperature are sensitive to models of the stellar atmosphere (see
Ref.~\cite{Suleimanov:2009dp} and references therein). Moreover, mass
and radius determination from X-ray burst (as in
Ref.~\cite{Ozel:2010fw}) may be hindered by systematic
uncertainties~\cite{Steiner:2010fz}.  For example Ozel, Baym, and
G\"uver inferred very small stellar radii from their recent analysis
of three X-ray bursts~\cite{Ozel:2010fw}, while Steiner, Lattimer, and
Brown extracted neutron star radii significantly larger (of the order
of 12 km)~\cite{Steiner:2010fz}. Still, we believe that such studies
will eventually become instrumental in constraining the dense matter
equation of state.

In the past, we have studied correlation between the neutron skin of
${}^{208}$Pb and the non-uniform solid crust of a neutron
star~\cite{Horowitz:2000xj}. For models with a stiff equation of
state, namely, one where the pressure increases rapidly with density,
it is energetically unfavorable to separate uniform nuclear matter
into regions of high and low densities. Thus, models with a stiff
equation of state predict low transition densities from non-uniform to
uniform neutron-rich matter. The neutron skin thickness in
${}^{208}$Pb also depends on the equation of state of neutron-rich
matter: the stiffer the equation of state the thicker the neutron
skin. Thus, an inverse relationship was established: {\sl the thicker
the neutron skin of a heavy nucleus, the lower the transition from
non-uniform to uniform neutron-rich matter. } This represents one of
the many examples that show the utility of constraining the neutron
skin thickness in ${}^{208}$Pb---and consequently the effective
interaction---for astrophysical applications.

Many relativistic effective interactions exist.  For example, the NL3
interaction~\cite{Lalazissis:1996rd,Lalazissis:1999} provides an
excellent description of the binding energy and charge radii of many
nuclei throughout the periodic table. When extrapolated to the
neutron-star domain, the NL3 effective interaction predicts a very
stiff equation of state that generates both large stellar radii and a
large maximum neutron star mass. On the other hand, the FSUGold (or
FSU for short) interaction~\cite{Todd-Rutel:2005fa} also provides a
good description of closed shell nuclei, but predicts a significantly
softer equation of state.  Indeed, whereas NL3 predicts a maximum
neutron star mass of 2.78 $M_\odot$, the FSU interaction predicts a
limiting mass almost one full solar mass smaller (of 1.72 $M_\odot$).
The softer equation of state emerges as one incorporates constraints
from giant resonances and heavy-ion collisions. At present, we are not
aware of any nuclear effective interaction that also incorporates
astrophysical constraints in the calibration procedure.  It is the aim
of the present manuscript to obtain a new effective interaction that
improves on both NL3 and FSU by incorporating some of the recent
constraints suggested by Steiner, Lattimer, and
Brown~\cite{Steiner:2010fz}. In particular, we will show that the EOS
predicted by the new model generates a limiting neutron star mass
intermediate between NL3 and FSU but stellar
radii smaller than both.

We have organized our article as follows.  In Section~\ref{reft} we
describe details of the fitting procedure and present results for
several ground-state properties of closed shell nuclei.  Next, we show
in Section~\ref{GR} how the new effective interaction reproduces the
properties of various nuclear collective modes. Predictions for the
equation of state and for various properties of neutron stars are
presented in Section~\ref{NS}. Finally, we conclude in Section
\ref{conclusions}.

\section{Relativistic Effective Field Theory}
\label{reft}
Our starting point will be the relativistic effective-field theory of
Ref.~\cite{Mueller:1996pm} supplemented with an isoscalar-isovector coupling as
introduced in Ref. \cite{Horowitz:2000xj}.  The interacting Lagrangian density
for this model is given by~\cite{Horowitz:2000xj,Mueller:1996pm}
\begin{eqnarray}
{\mathscr L}_{\rm int} &=&
\bar\psi \left[g_{\rm s}\phi   \!-\!
         \left(g_{\rm v}V_\mu  \!+\!
    \frac{g_{\rho}}{2}{\mbox{\boldmath $\tau$}}\cdot{\bf b}_{\mu}
                               \!+\!
    \frac{e}{2}(1\!+\!\tau_{3})A_{\mu}\right)\gamma^{\mu}
         \right]\psi \nonumber \\
                   &-&
    \frac{\kappa}{3!} (g_{\rm s}\phi)^3 \!-\!
    \frac{\lambda}{4!}(g_{\rm s}\phi)^4 \!+\!
    \frac{\zeta}{4!}   g_{\rm v}^4(V_{\mu}V^\mu)^2 +
   \Lambda_{\rm v} g_{\rho}^{2}\,{\bf b}_{\mu}\cdot{\bf b}^{\mu}
           g_{\rm v}^{2}V_{\nu}V^\nu\;.
 \label{LDensity}
\end{eqnarray}
The model contains an isodoublet nucleon field ($\psi$) interacting
via the exchange of two isoscalar mesons, the scalar sigma ($\phi$)
and the vector omega ($V^{\mu}$), one isovector meson, the rho (${\bf
b}^{\mu}$), and the photon ($A^{\mu}$). In addition to meson-nucleon
interactions, the Lagrangian density includes scalar and vector
self-interactions. (Note that while the original model allows for
$\rho$-meson self-interactions~\cite{Mueller:1996pm}, their
phenomenological impact has been documented to be small so they will
not be considered in this contribution). The scalar self-interaction
is responsible for reducing the compression modulus of nuclear matter
from the unrealistically large value of
$K\!=\!545$~MeV~\cite{Walecka:1974qa,Serot:1984ey} all the way down to
about $K\!=\!230$~MeV. This latter value appears to be consistent with
measurements of the isoscalar giant monopole resonance (GMR) in
${}^{208}$Pb~\cite{Youngblood:1999,Piekarewicz:2001nm,Piekarewicz:2002jd,
Colo:2004mj}.

Omega-meson self-interactions, as described by the parameter $\zeta$,
soften the equation of state at high density and can be tuned to
reproduce the maximum mass of a neutron star.  Indeed, M\"uller and
Serot found it was possible to build models with different values of $\zeta$
that reproduce the same observed properties at normal nuclear
densities, yet produced maximum neutron star masses that differ by
almost one solar mass~\cite{Mueller:1996pm}. In particular, models
with $\zeta\!=\!0$ predict large limiting masses of about $2.8
M_{\odot}$---even for models with a soft symmetry energy. In contrast,
the nonlinear coupling constant $\Lambda_{\rm v}$ was included to
modify the density-dependence of the symmetry energy.  Tuning
$\Lambda_{\rm v}$ provides a simple and efficient method of
softening the symmetry energy without compromising the success
of the model in reproducing well determined ground-state
observables~\cite{Todd-Rutel:2005fa}. Here we will adjust
$\Lambda_{\rm v}$ by assuming a relatively small value
for the neutron skin thickness of $^{208}$Pb.
Many properties of finite nuclei, such as the binding energies and
charge radii of closed shell nuclei, are insensitive to the values of
$\zeta$ and $\Lambda_{\rm v}$.  Indeed, the NL3 effective interaction
reproduces these observables without ever introducing these two
empirical parameters. Our approach here is to accommodate newly
proposed astrophysical constraints that suggest that, relative to the
FSU interaction, the equation of state of stellar matter must be
slightly softer at intermediate densities and stiffer at high
densities~\cite{Steiner:2010fz}.

%We then adopt the following simple procedure. We start from the
%FSU effective interaction and slightly tuned both $\zeta$ and
%$\Lambda_{\rm v}$ to obtain a larger limiting stellar mass and a
%smaller neutron skin thickness of $^{208}$Pb. Then, we
%first fix these
%two parameters, and then fit the other parameters to reproduce finite
%nuclei as we describe below.  Finally we adjust $\zeta$ to obtain a
%desired value for the neutron star maximum mass and $\Lambda_{\rm v}$
%to obtain a desired value for the neutron radius in $^{208}$Pb.

The new effective interaction is generated by adopting the following
procedure. We start from the FSU parameter
set~\cite{Todd-Rutel:2005fa} as listed in Table~\ref{Table1}.  This
parameter set reproduces ground-state properties and collective
excitations of closed shell nuclei. However, this interaction has a
relatively large value of $\zeta=0.06$. This reduces the pressure at
high densities and generates a relatively small neutron star maximum
mass of 1.72~$M_\odot$.  Note that, to date, the only terrestrial
constraint on the high-density component of the EOS comes from
energetic heavy-ion collisions~\cite{Danielewicz:2002pu}. The FSU
equation of state fits these data comfortable within the errors. Yet,
the reported errors are large enough to accommodate slightly stiffer
equations of state. As we aim for an interaction with a somewhat
larger maximum mass of about 2.0~$M_\odot$, we reduce the value of
$\zeta$ from 0.06 to 0.03.  Next, we refit the isoscalar parameters
($g_{\rm s}$, $g_{\rm v}$, $\kappa$, and $\lambda$) to maintain the
saturation properties of symmetric nuclear matter at their FSU values,
namely, a saturation density of 0.148 fm$^{-3}$ (or a Fermi momentum
of $k_{\rm F}\!=\!1.30$~fm$^{-1}$), an energy per nucleon of
$E/A\!=\!-16.30$ MeV, and an incompressibility coefficient of
$K\!=\!230$ MeV.

%%%%%%%%%%%%%%%%%%%%%%%%%%%%%%%%%%%%%%%%%%%%
\begin{widetext}
\begin{center}
\begin{table}[h]
\begin{tabular}{|l||c|c|c|c|c|c|c|c|c|c|}
 \hline
 Model & $m_{\rm s}$  & $m_{\rm v}$  & $m_{\rho}$
       & $g_{\rm s}^2$ & $g_{\rm v}^2$ & $g_{\rho}^2$
       & $\kappa$ & $\lambda$ & $\zeta$ & $\Lambda_{\rm v}$\\
 \hline
 \hline
 NL3       & 508.194 & 782.501 & 763.000 & 104.3871 & 165.5854 &  79.6000
               & 3.8599  & $-$0.015905 & 0.00 & 0.000 \\
 FSU        & 491.500 & 782.500 & 763.000 & 112.1996 & 204.5469 & 138.4701
               & 1.4203  & $+$0.023762 & 0.06 & 0.030 \\
 IU-FSU   & 491.500 & 782.500 & 763.000 &  99.4266 & 169.8349 & 184.6877
               & 3.3808  & $+$0.000296 & 0.03 & 0.046 \\
\hline
\end{tabular}
\caption{Parameter sets for the three models discussed in the text.
  The parameter $\kappa$ and the meson masses $m_{\rm s}$, $m_{\rm v}$,
  and $m_{\rho}$ are all given in MeV. The nucleon mass has been fixed at
  $M\!=\!939$~MeV in all the models.}
\label{Table1}
\end{table}
\end{center}
\end{widetext}
%%%%%%%%%%%%%%%%%%%%%%%%%%%%%%%%%%%%%%%%%%%%%%%%%%%%%%%%%%%%%%%%%

%%%%%%%%%%%%%%%%%%%%%%%%%%%%%%%%%%%%%%%%%%%%%%%%%%%%%%%%%%%%%%%%%
\begin{table}
\begin{tabular}{|l||c|c|c|c|c|}
 \hline
 Model & $\rho_{0}~({\rm fm}^{-3}) $ & $\varepsilon_{0}$~(MeV)
           & $K_{0}$~(MeV) & $J$~(MeV) & $L$~(MeV) \\
\hline
 \hline
 NL3    &  0.148  & $-$16.24 & 271.5 & 37.29 & 118.2 \\
 FSU     &  0.148  & $-$16.30 & 230.0 & 32.59 & 60.5  \\
IU-FSU & 0.155  & $-$16.40 & 231.2 & 31.30 & 47.2  \\

\hline
\end{tabular}
\caption{Bulk parameters characterizing the behavior of
              infinite nuclear matter at saturation density
              $\rho_{_{0}}$. The quantities $\varepsilon_{_{0}}$
              and $K_{0}$ represent the binding energy per
              nucleon and incompressibility coefficient of
              symmetric nuclear matter, whereas $J$ and
              $L$ represent the energy and slope of the
              symmetry energy at saturation density.}
\label{Table1b}
\end{table}
%%%%%%%%%%%%%%%%%%%%%%%%%%%%%%%%%%%%%%%%%%%%%%%%%%%%%%%%%%%%%%%%%

We then increase the isoscalar-isovector coupling constant
$\Lambda_{\rm v}$ from its FSU value of 0.03 to 0.046. This change
softens the density dependence of the symmetry energy and reduces the
neutron radius in $^{208}$Pb.  We aim for a modest value of the
neutron skin thickness in $^{208}$Pb of about
$R_{n}\!-\!R_{p}\!=\!0.16$ fm.  This value, although smaller as
compared to other relativistic mean field models, is close to those
predicted by many non-relativistic models.  Note that when we change
$\Lambda_{\rm v}$, we also change $g_\rho$ in order to maintain the
symmetry energy fixed at $\sim\!26$~MeV at a Fermi momentum of $k_{\rm
F}\!=\!1.15~{\rm fm}^{-1}$. This procedure and the justification for
it are described in detail in Ref.~\cite{Horowitz:2000xj}.  Increasing
$\Lambda_{\rm v}$ in this fashion reduces the density dependence of
the symmetry energy. This generates a symmetry energy at low densities
that is {\sl larger} than the one for FSU.  As a result, protons near
the surface of $^{208}$Pb are pulled closer to the neutrons in an
effort to minimize the neutron-proton asymmetry. This leads to an
increase in the charge radius of $^{208}$Pb.  In order to maintain the
charge radius in $^{208}$Pb at its experimental value we must slightly
increase the Fermi momentum of symmetric nuclear matter from 1.30
to 1.318 fm$^{-1}$. We then refit the parameters of the model to keep
the other saturation properties of nuclear matter intact. Finally, the
scalar coupling $g_{\rm s}$ is slightly tuned to improve the fit to
the binding energies of closed shell nuclei. Note that we have not
changed the scalar mass $m_{\rm s}$ from its FSU value.

\begin{figure}[htb]
\vspace{-0.05in}
\includegraphics[width=0.8\columnwidth,angle=0]{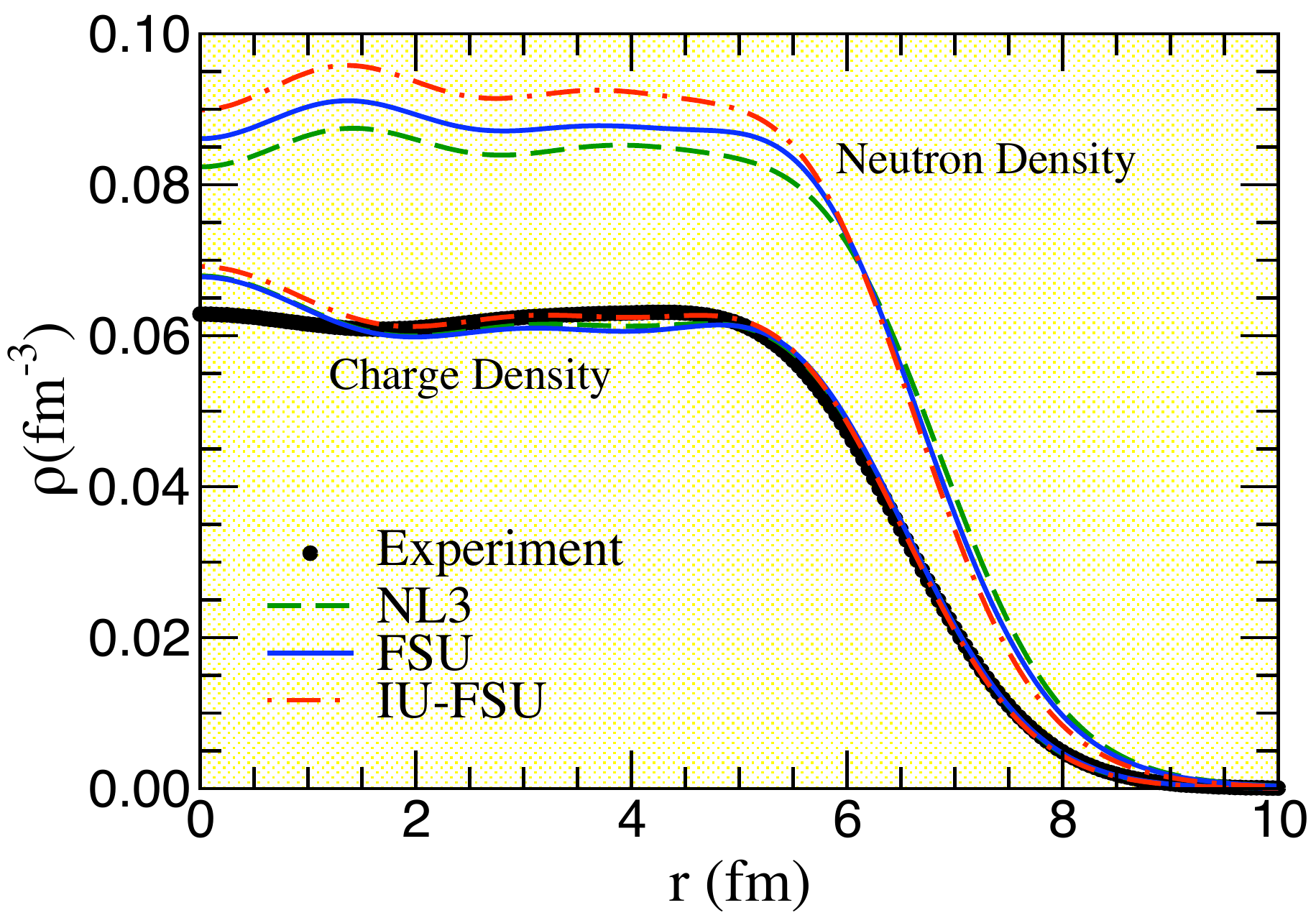}
\caption{(Color online) Model predictions for the charge and neutron densities of
                ${}^{208}$Pb. The experimental charge density is from
                Ref.~\cite{DeJager:1987qc}.}
\label{Fig1}
\end{figure}

The resulting parameter set---henceforth referred to as the {\sl
Indiana University-Florida State University (IU-FSU) interaction}---is
listed in Table~\ref{Table1}. The resulting bulk properties of
infinite nuclear matter have been collected in Table~\ref{Table1b}.
In addition, predictions for several ground-state properties of closed
shell nuclei are listed in Table \ref{Table2} and compared against
other theoretical models and experiment (when available).  Finally,
predictions for the charge and neutron densities of $^{208}$Pb are
displayed in Fig.~\ref{Fig1}.  Whereas significant differences can be
observed in the prediction of the neutron densities, the difference
among the models is small for the charge density.  Note that although
the parameters of the model will eventually be determined from an
accurate calibration procedure, the new model reproduces rather well the
charge radii and binding energies of closed shell nuclei. However, the
IU-FSU interaction predicts the rather modest value of
$R_n\!-\!R_p\!=\!0.16$ fm for the, as yet unknown, neutron skin thickness
in $^{208}$Pb.

%%%%%%%%%%%%%%%%%%%%%%%%%%%%%%%%%%%%%%%%%%%%%%%%%%%%%%%%%%%%%%%%%
  \begin{table}[h]
  \begin{tabular}{|c|c|c|c|c|c|}
    \hline
    Nucleus & Observable & Experiment & NL3 & FSU & IU-FSU \\
    \hline
    \hline
    ${}^{40}$Ca & $B/A$~(MeV)  & $8.55$ & $\phantom{-}8.54$ & $\phantom{-}8.54$  & $\phantom{-}8.53$ \\
     & $R_{\rm ch}$~(fm) & $3.45$ & $\phantom{-}3.46$ & $\phantom{-}3.42$  & $\phantom{-}3.41$  \\
     & $R_{n}\!-\!R_{p}$~(fm)  &  ---  & $-0.05$ & $-0.05$  & $-0.05$ \\
   \hline
    ${}^{48}$Ca & $B/A$~(MeV)  & $8.67$ & $\phantom{-}8.64$ & $\phantom{-}8.58$ & $\phantom{-}8.55$  \\
     & $R_{\rm ch}$~(fm) & $3.45$ & $\phantom{-}3.46$ & $\phantom{-}3.45$ & $\phantom{-}3.44$  \\
     & $R_{n}\!-\!R_{p}$~(fm) &   ---  & $\phantom{-}0.23$ & $\phantom{-}0.20$ & $\phantom{-}0.17$  \\
    \hline
    ${}^{90}$Zr & $B/A$~(MeV) & $8.71$ & $\phantom{-}8.69$ & $\phantom{-}8.68$ & $\phantom{-}8.67$ \\
     & $R_{\rm ch}$~(fm)  & $4.26$ & $\phantom{-}4.26$ & $\phantom{-}4.25$ & $\phantom{-}4.23$  \\
     & $R_{n}\!-\!R_{p}$~(fm) &   ---  & $\phantom{-}0.11$ & $\phantom{-}0.09$ & $\phantom{-}0.07$  \\
    \hline
   ${}^{132}$Sn & $B/A$~(MeV) & $8.36$ & $\phantom{-}8.37$ & $\phantom{-}8.34$ & $\phantom{-}8.33$ \\
     & $R_{\rm ch}$~(fm)   &   ---  & $\phantom{-}4.70$ & $\phantom{-}4.71$ & $\phantom{-}4.68$  \\
     & $R_{n}\!-\!R_{p}$~(fm) &   ---  & $\phantom{-}0.35$ & $\phantom{-}0.27$ & $\phantom{-}0.22$  \\
    \hline
   ${}^{208}$Pb & $B/A$~(MeV)  & $7.87$ & $\phantom{-}7.88$ & $\phantom{-}7.89$  & $\phantom{-}7.89$  \\
     & $R_{\rm ch}$~(fm)   & $5.50$ & $\phantom{-}5.51$ & $\phantom{-}5.52$ & $\phantom{-}5.48$  \\
     & $R_{n}\!-\!R_{p}$~(fm) &   ---  & $\phantom{-}0.28$ & $\phantom{-}0.21$ & $\phantom{-}0.16$  \\
    \hline
  \end{tabular}
 \caption{Experimental data for the binding energy per nucleon and
               charge radii for several doubly magic nuclei. Results are
               presented for the three models employed in the text alongside
               their predictions for the neutron skin thickness of these nuclei.}
 \label{Table2}
 \end{table}
%%%%%%%%%%%%%%%%%%%%%%%%%%%%%%%%%%%%%%%%%%%%%%%%%%%%%%%%%%%%%%%%%

%%%%%%%%%%%%%%%%%%%%%%%%%%%%%%%%%%%%%%%%%%%%%%%%%%%%%%%%%%%%%%%%%
%%%%%%%%%%%%%%%%%%%%%%%%%%%%%%%%%%%%%%%%%%%%%%%%%%%%%%%%%%%%%%%%%

\section{Giant Resonances}
\label{GR}

At the time that the FSU effective interaction was calibrated, the
only evidence in favor of a symmetry energy softer than NL3 came from
nuclear collective excitations, particularly the mass dependence of
the isoscalar giant monopole resonance (GMR).  Whereas NL3---with
an incompressibility coefficient of $K\!=271\!$~MeV---was able to
reproduce the centroid energy of the GMR in ${}^{208}$Pb, it
overestimated the GMR in ${}^{90}$Zr. It was later confirmed that the
success of the NL3 interaction in reproducing the breathing mode in
${}^{208}$Pb was accidental, as it resulted from a combination of both
a stiff EOS for symmetric nuclear matter and a stiff symmetry
energy~\cite{Piekarewicz:2003br}.  Indeed, with a relatively large
neutron-proton asymmetry of $\alpha\!\equiv\!(N-Z)/A=0.212$, the GMR
in ${}^{208}$Pb probes the incompressibility of neutron-rich matter,
rather than that of symmetric matter. Given that the incompressibility
of neutron-rich matter softens with
$\alpha$~\cite{Piekarewicz:2008nh}, NL3 could reproduce the GMR in
${}^{208}$Pb by cancelling its stiff incompressibility coefficient $K$
with a correspondingly stiff symmetry energy.  This cancellation,
however, is incomplete in ${}^{90}$Zr because its nucleon asymmetry is
almost twice as small as that of ${}^{208}$Pb. Thus, to develop the
FSU effective interaction we used the GMR in $^{90}$Zr, rather than in
${}^{208}$Pb, to fix the incompressibility coefficient of symmetric
nuclear matter. Having fixed $K$, we used the GMR---and the isovector
giant dipole resonance (IVGDR)---in ${}^{208}$Pb to constrain the
density dependence of the symmetry energy~\cite{Todd-Rutel:2005fa}.
As we shall see below (see Table~\ref{Table3}), the newly proposed
IU-FSU effective interaction continues to reproduce the centroid
energy of these three modes.

The distribution of both isoscalar monopole and isovector dipole
strength were computed in a relativistic random-phase approximation
(RPA).  The first step in calculating the RPA response is the
computation of the mean-field ground state in a self-consistent
mean-field approximation. Once self-consistency is achieved, three
important pieces of information become available: (i) the
single-particle energies of the occupied orbitals, (ii) their
single-particle wave functions, and (iii) the self-consistent
mean-field potential.  This mean-field potential---without any
modification---must then be used to generate the nucleon propagator;
only then can one ensure the conservation of the vector current. The
nucleon propagator, which is an essential building block of the
uncorrelated polarization tensor, is computed non-spectrally to avoid
any dependence on the artificial cutoffs and truncations that plague
most spectral approaches. Moreover, a great merit of the non-spectral
approach is that the continuum is treated exactly. Once the
uncorrelated polarization insertion is computed, the correlated RPA
polarization is obtained by iterating it to all orders with the
consistent residual interaction. The distribution of strength is then
obtained directly from the imaginary part of the polarization
tensor. A detailed description of the relativistic RPA approach may be
found in Ref.~\cite{Piekarewicz:2000nm,Piekarewicz:2001nm}.
Results for the centroid energies of the giant monopole and dipole
resonances are presented in Table \ref{Table3} for the three effective
interactions considered in the text. Unlike the NL3 model, the results
obtained with both the FSU and the IU-FSU effective interactions are
consistent with experiment.

%%%%%%%%%%%%%%%%%%%%%%%%%%%%%%%%%%%%%%%%%%%%%%%%%%%%%%%%%%%%%%%%%
  \begin{table}[h]
  \begin{tabular}{|c|c|c|c|c|c|}
    \hline
    Nucleus & Observable & Experiment & NL3 & FSU & IU-FSU \\
    \hline
    \hline
    ${}^{208}$Pb & GMR   (MeV) & $14.17\pm0.28$ & $14.32$ & $14.04$  & $14.17$ \\
    ${}^{90}$Zr  & GMR   (MeV) & $17.89\pm0.20$ & $18.62$ & $17.98$   & $17.87$\\
    ${}^{208}$Pb & IVGDR (MeV) & $13.30\pm0.10$ & $12.70$ & $13.07$ & $13.24$\\
    \hline
  \end{tabular}
 \caption{Centroid energies for the GMR in ${}^{208}$Pb and
          ${}^{90}$Zr, and the peak energy for the IVGDR in
          ${}^{208}$Pb.  Experimental data are extracted from
          Refs.~\cite{Youngblood:1999} and~\cite{Rit93_PRL70}.}
  \label{Table3}
 \end{table}
%%%%%%%%%%%%%%%%%%%%%%%%%%%%%%%%%%%%%%%%%%%%%%%%%%%%%%%%%%%%%%%%%

\section{Equation of State and Neutron Star Structure}
\label{NS}

\begin{figure}[h]
\vspace{-0.05in}
\includegraphics[width=0.8\columnwidth,angle=0]{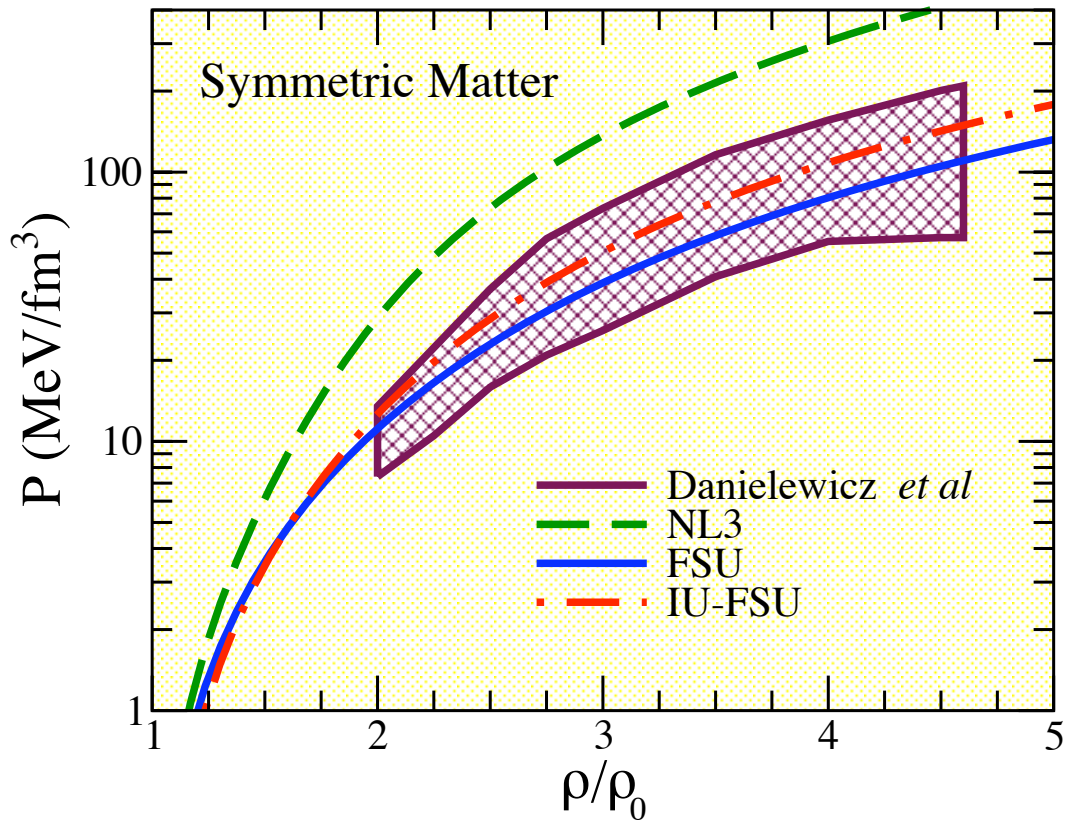}
\caption{(Color online) The equation of state---pressure $P$ {\sl vs}
baryon density---of symmetric nuclear matter. Here $\rho_{0}$ is the
density of nuclear matter at saturation and the shaded area represents
the EOS extracted from the analysis of Ref.~\cite{Danielewicz:2002pu}.}
\label{Fig2}
\end{figure}

Although the full complexity of the quantum system can not be tackled
exactly, the energy density of asymmetric nuclear matter may be
computed in a {\sl mean-field} (MF) approximation. In the MF
approximation all the meson fields are replaced by their classical
expectation values and their solution can be readily obtained by
solving the classical Euler-Lagrange equations of motion. The sole
remnant of quantum behavior is in the treatment of the nucleon field
which emerges from a solution to the Dirac equation in the presence of
appropriate scalar and vector
potentials~\cite{Serot:1984ey,Serot:1997xg}.  Following standard
mean-field practices, the energy density of the system is given by the
following expression:
\begin{eqnarray}
{\mathscr E}(\rho,\alpha)  &=&
 \frac{1}{\pi^{2}}\int_{0}^{k_{\rm F}^{p}} k^{2}\sqrt{k^2+M^{*2}}\,dk +
 \frac{1}{\pi^{2}}\int_{0}^{k_{\rm F}^{n}} k^{2}\sqrt{k^2+M^{*2}}\,dk +
 \frac{1}{2}\left(\frac{m_{\rm s}^{2}}{g_{\rm s}^{2}}\right)\Phi_{0}^{2} +
 \frac{\kappa}{6}\Phi_{0}^{3} + \frac{\lambda}{24}\Phi_{0}^{4}
 \nonumber \\ &+&
 \frac{1}{2}\left(\frac{m_{\rm v}^{2}}{g_{\rm v}^{2}}\right)W_{0}^{2}+
 \frac{\zeta}{8}W_{0}^{4} +
 \frac{1}{2}\left(\frac{m_{\rho}^{2}}{g_{\rho}^{2}}\right)B_{0}^2+
 3\Lambda_{\rm v}W_{0}^{2} B_{0}^{2} \;,
\label{EDensity2}
\end{eqnarray}
where $\rho$ is the baryon density of the system and
$\alpha\!=\!(\rho_{n}\!-\!\rho_{p})/\rho$ is the neutron-proton
asymmetry. Alternatively, the energy density may be written in terms
of the proton and neutron Fermi momenta $k_{\rm F}^{p}$ and $k_{\rm
F}^{n}$.  Note that the following definitions have been introduced:
$\Phi_{0}\!\equiv\!g_{\rm s}\phi_{0}$, $W_{0}\!\equiv\!g_{\rm
v}V_{0}$, and $B_{0}\!\equiv\!g_{\rho}b_{0}$.  Also note that since
the MF approximation is thermodynamically consistent, the pressure of
the system (at zero temperature) may be obtained either from the
energy-momentum tensor or from the energy density and its first
derivative. That is,
\begin{equation}
 P(\rho,\alpha)=\rho\frac{\partial{\mathscr E}}{\partial\rho}
            -{\mathscr E} \;.
\label{Pressure}
\end{equation}
The equation of state (pressure as a function of baryon density) for
symmetric nuclear matter ($k_{\rm F}^{p}\!=\!k_{\rm F}^{n}$ and
$B_{0}\!\equiv\!0$) is displayed in Fig~\ref{Fig2}. By design, the FSU
and IU-FSU interactions have (almost) the same incompressibility
coefficient so they predict similar pressures at low to intermediate
densities.  However, at higher densities the IU-FSU interaction is
stiffer because it has a smaller value of $\zeta$ as compared to the
FSU interaction.  To date, the only terrestrial constraint on the
high-density component of the EOS comes from energetic heavy-ion
collisions.  Both of these models predict equations of state that are
consistent with the phenomenological flow analysis by Danielewicz,
Lacey, and Lynch~\cite{Danielewicz:2002pu}. In contrast, the NL3
interaction, although enormously successful in reproducing
ground-state properties of finite nuclei, predicts an equation of
state that is significantly stiffer than the phenomenological extraction.

%%%%%%%%%%%%%%%%%%%%%%%%%%%%%%%%%%%%%%%%%%%%%%%%%%%%%%%%%%%%%%%%%
%%%%%%%%%%%%%%%%%%%%%%%%%%%%%%%%%%%%%%%%%%%%%%%%%%%%%%%%%%%%%%%%%

\begin{figure}[h]
\vspace{-0.05in}
\includegraphics[width=0.8\columnwidth,angle=0]{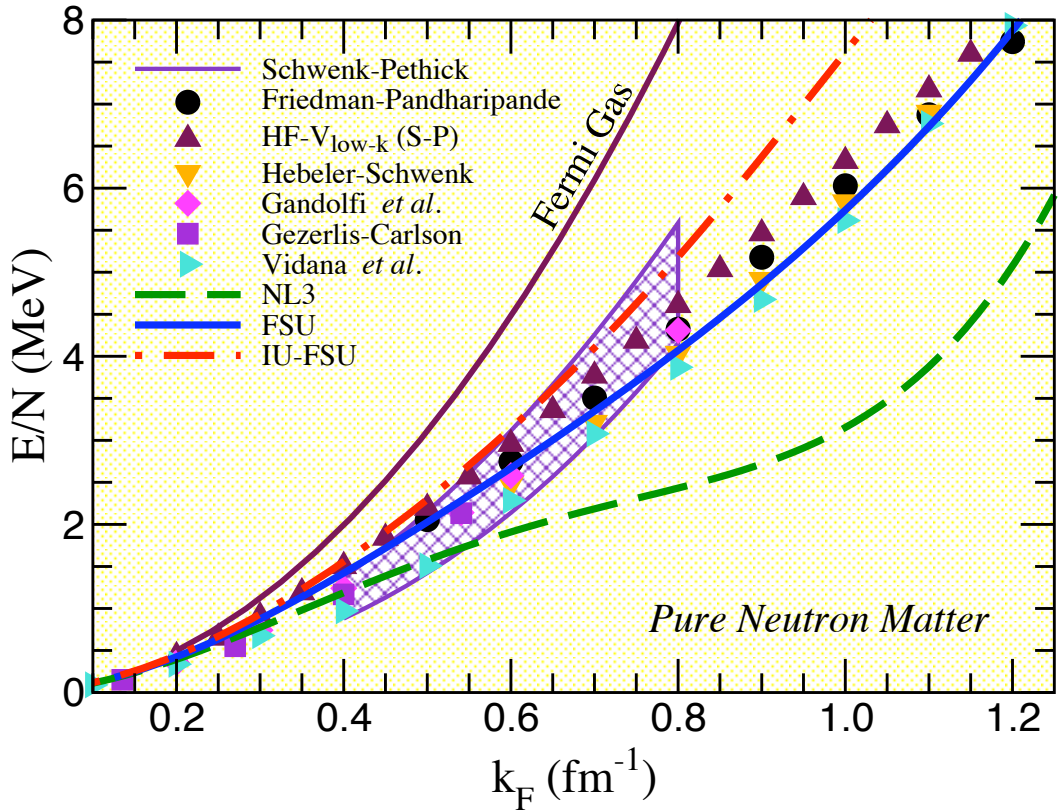}
\caption{(Color online) Energy per nucleon as a function of the
 Fermi momentum for pure neutron matter.}
\label{Fig3}
\end{figure}

The equation of state of pure neutron matter (PNM) provides a
powerful theoretical constraint on models of the effective
interaction. By building on the universal behavior of dilute Fermi
gases with an infinite scattering length, significant progress has
been made in constraining the equation of state of pure neutron
matter~\cite{Schwenk:2005ka}.  To date, a variety of models using
different neutron-neutron interactions and a variety of many-body
techniques have been employed to compute the EOS of dilute neutron
matter (see Ref.~\cite{Gezerlis:2009iw} and references therein). In
Fig.~\ref{Fig3} we display the energy per particle of pure neutron
matter for a variety of microscopic
approaches~\cite{Gezerlis:2009iw} and for the three relativistic
effective interactions discussed in the text.  In most relativistic
descriptions the isovector interaction is modelled by the exchange
of a single vector-isovector (``rho'') meson with its (Yukawa)
coupling to the nucleon tuned to reproduce the symmetry energy at
saturation density.  Invariably, such a simple prescription
generates a stiff symmetry energy. In turn, this yields an EOS for
dilute neutron matter that is inconsistent with the
model-independent results of Schwenk and Pethick (denoted by the
shaded area in Fig.~\ref{Fig3}). The FSU isovector interaction
improves on NL3 by adding an isoscalar-isovector mixing term
($\Lambda_{\rm v}$) that softens the symmetry energy. Recall that
such a softening is required to reproduce the centroid energies of
various collective modes (see Table~\ref{Table3}). Without any
further adjustment, the FSU interaction is also consistent with the
theoretical constraints (see solid blue line). Finally, the IU-FSU
interaction---with an additional softening relative to FSU---remains
within the theoretical ``error bars''.

%%%%%%%%%%%%%%%%%%%%%%%%%%%%%%%%%%%%%%%%%%%%%%%%%%%%%%%%%%%%%%%%%
%%%%%%%%%%%%%%%%%%%%%%%%%%%%%%%%%%%%%%%%%%%%%%%%%%%%%%%%%%%%%%%%%

We conclude this section by computing the {\sl mass-vs-radius} ({\sl
M-R}) relationship of neutron stars. Given that the structure of
neutron stars is sensitive to the EOS over an enormous range of
densities, we must resort to various descriptions.  For the
non-uniform outer crust we employ the equation of state of Baym,
Pethick, and Sutherland~\cite{Baym:1971pw}.  At densities of about a
third of nuclear-matter saturation density, uniformity in the system
is restored. For this (liquid-core) region we use an EOS generated
from the relativistic models discussed in the text.  Although
significant progress has been made in simulating the exotic structure
of the inner crust ({\sl i.e.,} the region between the outer crust and
the liquid core) a detailed equation of state is still missing in this
region.  Hence, we resort to a fairly accurate polytropic EOS to
interpolate between the solid crust and the uniform liquid
core~\cite{Link:1999ca,Carriere:2002bx}. The transition from the
uniform liquid core to the solid crust is sensitive to the density
dependence of the symmetry energy. The crust-to-core transition
density $\rho_{t}$ is obtained from an RPA stability analysis to
determine the onset of the instability to small amplitude density
oscillations~\cite{Carriere:2002bx}. With such an equation of state in
hand, we can now proceed to integrate the Tolman-Oppenheimer-Volkoff
equations to determine the structure of neutron stars.

\begin{figure}[h]
\vspace{-0.05in}
\includegraphics[width=0.8\columnwidth,angle=0]{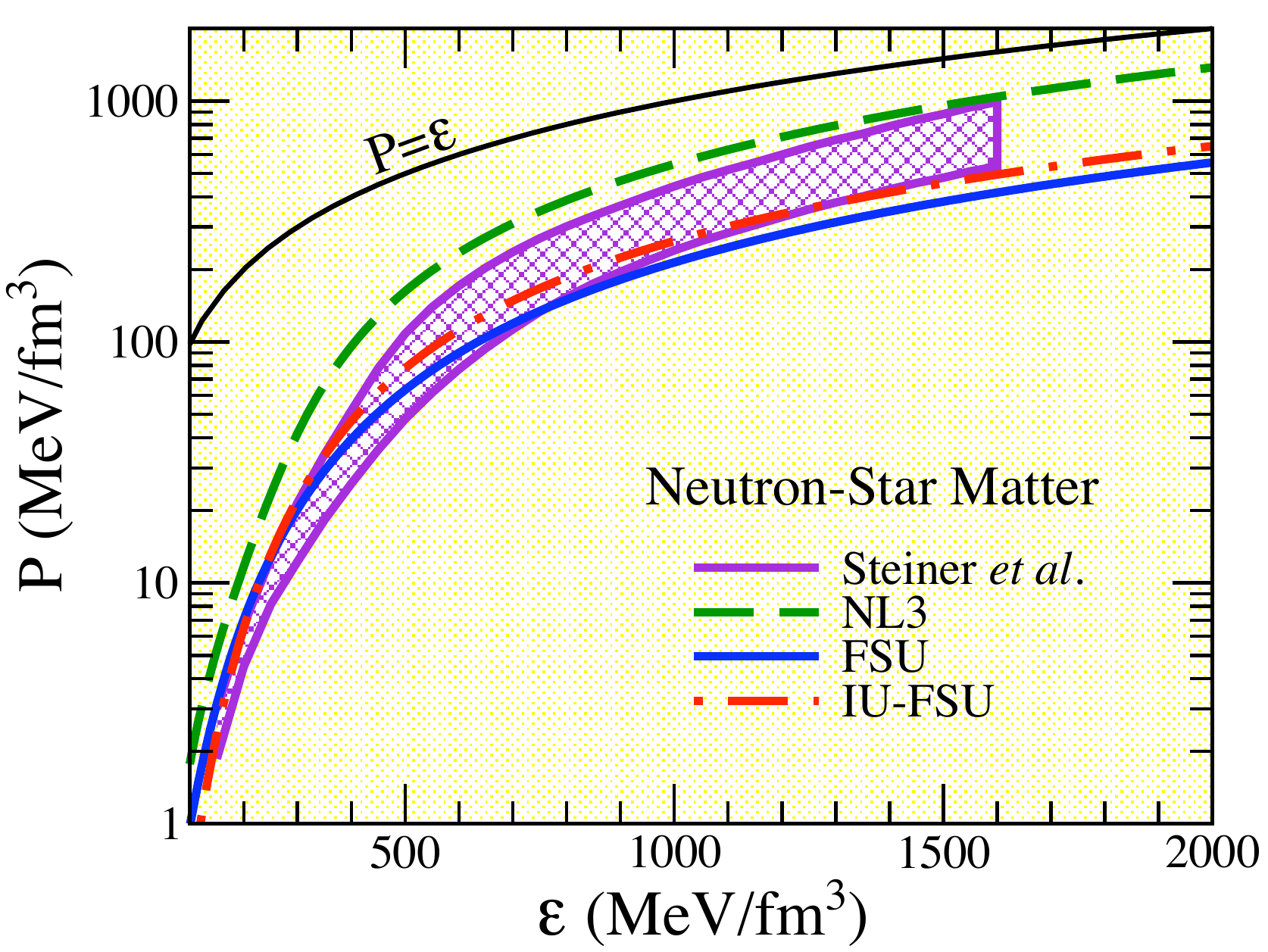}
\caption{(Color online) Equation of state---pressure {\sl vs} energy
density---of neutron-rich matter in beta equilibrium. The shaded region
displays the observational constraint extracted from Ref.~\cite{Steiner:2010fz}.
The solid black line ($P\!=\!{\mathcal E}$) denotes the stiffest
possible equation of state consistent with causality.}
\label{Fig4}
\end{figure}

In Fig.~\ref{Fig4} we display predictions for the equation of state of
stellar matter ({\sl i.e.,} neutron-rich matter in beta equilibrium)
for the three models discussed in the text, alongside the constraint
extracted from Ref.~\cite{Steiner:2010fz}. For reference, also shown is the
stiffest ($P\!=\!{\mathcal E}$) possible equation of state consistent
with causality. As suggested in Ref.~\cite{Steiner:2010fz}, the FSU interaction
appears slightly stiff at intermediate densities and too soft at high
densities. The IU-FSU interaction successfully corrects both of these
problems. Note, however, that the observational EOS can accommodate
even stiffer equations of state. We stress that, if required, this can
be implemented rather easily by reducing even further the value
of $\zeta$.

%%%%%%%%%%%%%%%%%%%%%%%%%%%%%%%%%%%%%%%%%%%%%%%%%%%%%%%%%%%%%%%%%
%%%%%%%%%%%%%%%%%%%%%%%%%%%%%%%%%%%%%%%%%%%%%%%%%%%%%%%%%%%%%%%%%

\begin{figure}[h]
\vspace{-0.05in}
\includegraphics[width=0.8\columnwidth,angle=0]{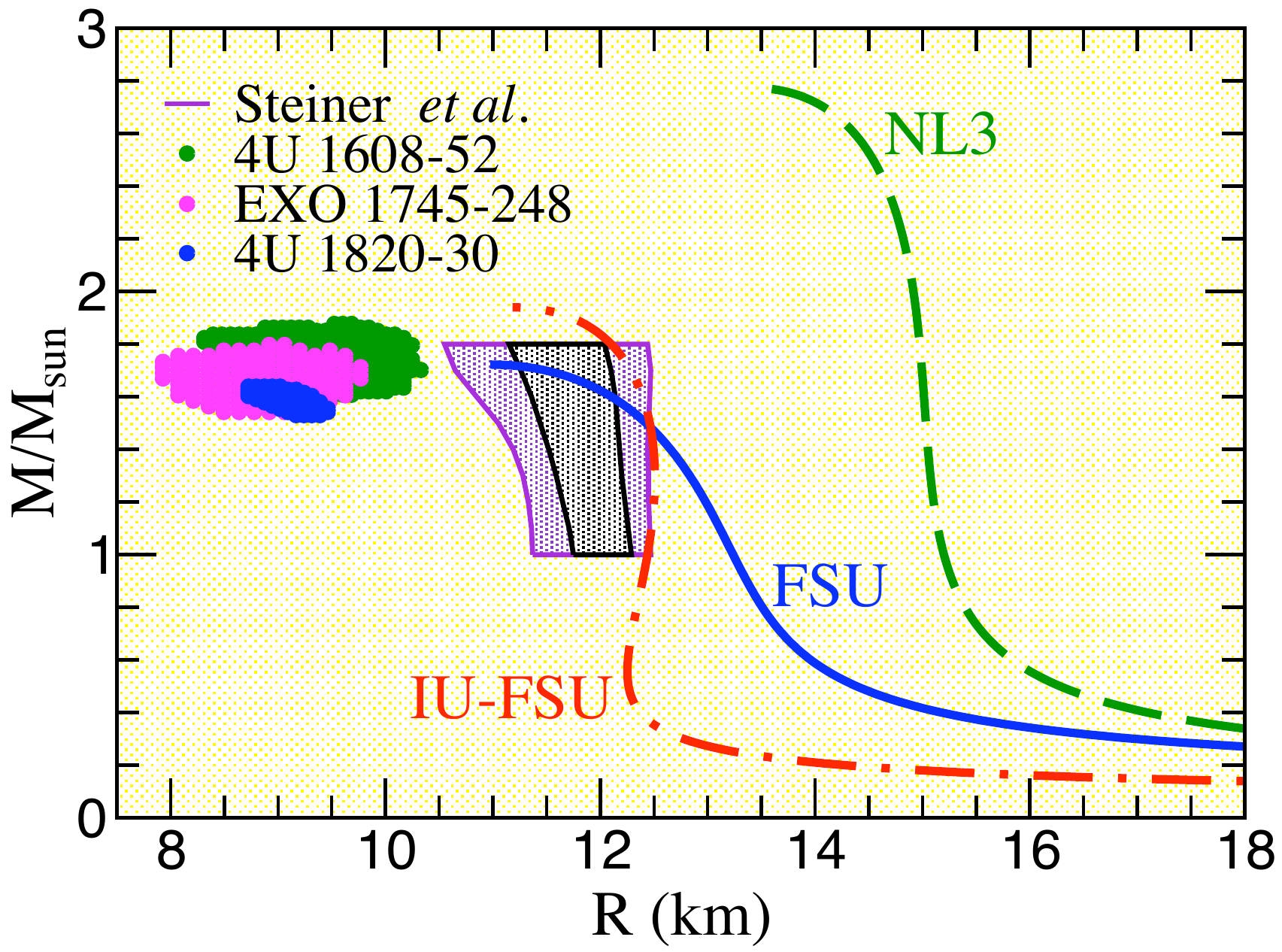}
\caption{(Color online) {\sl Mass-vs-Radius} relation predicted by
 the three relativistic mean-field models discussed in the text. The
 observational data that suggest very small stellar radii represent
1$\sigma$ confidence contours for the three neutron stars reported
in Ref.~\cite{Ozel:2010fw}. The two shaded areas that suggest larger radii
are 1$\sigma$ and 2$\sigma$ contours extracted from the analysis of
Ref.~\cite{Steiner:2010fz}.}
\label{Fig5}
\end{figure}

In Fig.~\ref{Fig5} we display predictions for the {\sl M-R} relation
and compare them against observational constraints extracted from the
analyses of Refs.~\cite{Ozel:2010fw} and~\cite{Steiner:2010fz}. The
very stiff behavior of the EOS predicted by NL3 is immediately ruled
out by both observational constraints.  In the case of the FSU
effective interaction, we recently identified a conflict when compared
with the analysis by Ozel and collaborators~\cite{Fattoyev:2010rx}.
Whereas the FSU model seems to generate an EOS that is consistent with
observation, the predicted stellar radii were simply too
large. Relative to the analysis by Steiner Lattimer, and
Brown~\cite{Steiner:2010fz}, the FSU predictions overestimate the
stellar radius---although not as much as suggested
by~\cite{Ozel:2010fw}---but underestimates the maximum neutron star
mass. The IU-FSU interaction, with a softer EOS at intermediate
densities and a stiffer one at high densities, is motivated in
response to these findings. Indeed, the IU-FSU interaction predicts a
maximum stellar mass of 1.94~$M_\odot$ and a stellar radius of
$R=12.49$ km for a 1.4~$M_\odot$ neutron star (see
Table~\ref{Table4}).  These predictions are consistent with the
$2\sigma$ values extracted from X-ray observations by
Steiner, Lattimer, and Brown~\cite{Steiner:2010fz}. However, they are
well outside the limits extracted by Ozel and
collaborators~\cite{Ozel:2010fw}.  Given that such extractions depend
critically on the models used to simulate X-ray bursts, much work
remains to be done to reconcile these two analyses. We close this
section by listing in Table~\ref{Table4} predictions for several
important neutron-star properties.  In addition to the properties
already discussed (such as masses and radii), the table includes the
minimum density required for the onset of the direct Urca process and
the minimum stellar mass that may cool down by the direct Urca
process. Note that small neutron star radii and enhanced cooling are
generally regarded as good indicators of a possible phase transition
in the stellar core.

%%%%%%%%%%%%%%%%%%%%%%%%%%%%%%%%%%%%%%%%%%%%%%%%%%%%%%%%%%%%%%%%%
%%%%%%%%%%%%%%%%%%%%%%%%%%%%%%%%%%%%%%%%%%%%%%%%%%%%%%%%%%%%%%%%%

 %%%%%%%%%%%%%%%%%%%%%%%%%%%%%%%%%%%%%%%%%%%%%%%%%%%%%%%%%%%%%%%%%
  \begin{table}[h]
  \begin{tabular}{|c|c|c|c|}
    \hline
     Observable & NL3 & FSU & IU-FSU \\
    \hline
    \hline
     $\rho_{t}$~(fm$^{-3}$)              & $0.052$ & $0.076$ & $0.087$\\
     $R$~(km)                                 & $15.05$ & $12.66$ & $12.49$ \\
     $M_{\rm max}(M_{\odot})$             & $2.78$ & $1.72$ & $1.94$\\
     $\rho_{_{\rm Urca}}$~(fm$^{-3}$) & $0.21$ & $0.47$  & $0.61$  \\
     $M_{\rm Urca}(M_{\odot})$            & $0.84$ & $1.30$  & $1.77$  \\
%     $\Delta M_{\rm Urca}$                & $0.38$ & $0.06$  & $0.00$  \\
    \hline
  \end{tabular}
 \caption{Predictions for a few neutron-star observables. The various
          quantities are as follows: $\rho_{t}$ is the transition
          density from non-uniform to uniform neutron-rich
          matter, $R$ is the radius of a 1.4 solar-mass neutron star,
          $M_{\rm max}$ is the maximum mass, $\rho_{_{\rm Urca}}$ is
      the threshold density for the direct Urca process, and $M_{\rm Urca}$
          is the minimum mass neutron star that may cool down by the
          direct Urca process.}
  \label{Table4}
 \end{table}
%%%%%%%%%%%%%%%%%%%%%%%%%%%%%%%%%%%%%%%%%%%%%%%%%%%%%%%%%%%%%%%%%

\section{Conclusions}
\label{conclusions}

The structure of neutron stars depends sensitively on the equation of
state of neutron-rich matter in beta equilibrium. As such, nuclear
effective interactions play a critical role in astrophysical
applications. Given the large extrapolations in isospin and density
that are required to compute the EOS, it is imperative to constrain
the effective interaction.  Motivated by two recent analyses that have
provided simultaneous mass and radius information on neutron
stars~\cite{Ozel:2010fw,Steiner:2010fz}, we have introduced a new
effective interaction constrained by the ground state properties of
finite nuclei, their collective excitations, the properties of both
dilute and dense matter, and the structure of neutron stars. Thus,
experimental, theoretical, and observational constraints have all been
incorporated into the calibration procedure.

Following the conclusions of Ref.~\cite{Steiner:2010fz}---that suggest that the
accurately calibrated FSU effective interaction is slightly stiff at
intermediate densities but soft at high densities---we have used two
empirical parameters to correct these shortcomings. The new effective
interaction---dubbed {\sl ``IU-FSU''}---softens the EOS at
intermediate densities by reducing the neutron skin thickness of
$^{208}$Pb and stiffens the EOS at high density by increasing the
maximum neutron star mass relative to the FSU predictions.

Existing relativistic mean-field interactions are flexible enough to
reproduce, not only the charge density and binding energy of closed
shell nuclei, but also the neutron skin thickness of $^{208}$Pb and
the maximum neutron star mass~\cite{Fattoyev:2010rx,Fattoyev:2010tb}.
In particular, the density dependence of the symmetry energy is highly
sensitive to the empirical parameter $\Lambda_{\rm v}$. Thus, an
increase in the value of $\Lambda_{\rm v}$ softens the EOS at
intermediate densities and generates a neutron skin thickness in
$^{208}$Pb of only $R_n\!-\!R_p\!=\!0.16$~fm. For comparison, the NL3
and FSU effective interactions predict significantly larger neutron
skins: 0.28~fm and 0.21~fm, respectively.  To stiffen the EOS at high
densities we rely on the empirical parameter
$\zeta$~\cite{Mueller:1996pm,Fattoyev:2010rx,Fattoyev:2010tb}.
Including $\zeta$ may be used to change the maximum neutron star mass
by almost one solar mass without adversely affecting the bulk
properties of nuclear matter around saturation density. Relative to
the FSU model, the new IU-FSU interaction increases the maximum
neutron star mass from 1.72~$M_\odot$ to 1.94~$M_\odot$.

As more accurate data becomes available, the coupling
constants of the theory may need to be re-adjusted. The calibration
procedure outlined here can accomplish this task rapidly and
efficiently. Ultimately, however, one would rely on an {\sl accurately}
calibration procedure to determine the best-fit parameters of the
theory and to obtain correlations among the various observables.
Nevertheless, as it stands now the new IU-FSU relativistic effective
interaction reproduces: (a) the binding energies and charge radii of
closed-shell nuclei, (b) various nuclear giant (monopole and dipole)
resonances, (c) the low-density behavior of pure neutron matter, (d)
the high-density behavior of symmetric nuclear matter, and (e) the
mass-radius relation of neutron stars. In the
future, we plan to use the IU-FSU interaction to generate
a finite-temperature/finite-density equation of state for use in
supernova and neutron-star merger
simulations~\cite{Shen:2010pu,Shen:2010jd}.

\smallskip
This work was supported in part by DOE grants DE-FG02-87ER40365
and DE-FG05-92ER40750.

\bibliography{../ReferencesJP.bib}
\end{document}